\documentclass[acmsmall]{acmart}

\AtBeginDocument{%
  \providecommand\BibTeX{{%
    \normalfont B\kern-0.5em{\scshape i\kern-0.25em b}\kern-0.8em\TeX}}}

\copyrightyear{2024}
\acmYear{2024}
\setcopyright{rightsretained}
\acmConference[TEI '24]{Eighteenth International Conference on Tangible, Embedded, and Embodied Interaction}{February 11--14, 2024}{Cork, Ireland}
\acmBooktitle{Eighteenth International Conference on Tangible, Embedded, and Embodied Interaction (TEI '24), February 11--14, 2024, Cork, Ireland}
\acmDOI{10.1145/3623509.3635866}
\acmISBN{979-8-4007-0402-4/24/02}

\begin{document}

\title{Ichiyo: Fragile and Transient Interaction in Neighborhood}

\author{Hirofumi Shibata}
\authornote{All authors contributed equally to this research.}
\email{67-shibata-hirofumi@g.ecc.u-tokyo.ac.jp}
\affiliation{%
  \institution{The University of Tokyo}
  \city{Bunkyo}
  \country{Japan}
}

\author{Ayako Yogo}
\authornotemark[1]
\email{ayakorn45@gmail.com}
\affiliation{%
  \institution{The University of Tokyo}
  \city{Bunkyo}
  \country{Japan}
}

\author{Naoto Nishida}
\authornotemark[1]
\email{nawta@g.ecc.u-tokyo.ac.jp}
\orcid{0000-0001-9966-4664}
\affiliation{%
  \institution{The University of Tokyo}
  \city{Bunkyo}
  \country{Japan}
}

\author{Yu Shimada}
\authornotemark[1]
\email{y-shimada@g.ecc.u-tokyo.ac.jp}
\affiliation{%
  \institution{The University of Tokyo}
  \city{Bunkyo}
  \country{Japan}
}

\author{Toma Ishii}
\authornotemark[1]
\email{mutedblue2k20@gmail.com}
\affiliation{%
  \institution{The University of Tokyo}
  \city{Bunkyo}
  \country{Japan}
}


\renewcommand{\shortauthors}{Shibata, et al.}

\begin{abstract}
As the Internet develops, social networking and other communication tools have transformed people's relationships into something fast, visible, and geographically huge. 
However, these communication tools have not expanded opportunities for acquainting oneself with neighbors outside one's social network;
rather, they have comparatively diminished occasions for interacting with unfamiliar neighbors by prioritizing communication with existing friends.
Therefore, we invented the medium {\it Ichiyo} to increase the opportunities to think of neighbors walking along the same street or in the same neighborhood and to expand the imagination of those who pass by and those who used to be there. 
Thus, users can engage in indirect interaction.
We used commercially available laser cutters to engrave QR codes on leaves that are naturally found in our living space to prevent environmental invasion.
The QR codes lead to a communal space on the web where users can freely leave messages.
By engraving QR codes, information can be virtually expanded to be presented.
To get the feedback of {\it Ichiyo}, we let a total of several thousand people experience a new way of communication as a part of the exhibition ``iii Exhibition 2022'', an art exhibition at the University of Tokyo. 
A total of more than 1,000 leaves engraved with QR codes were prepared and scattered at the exhibition site and along the road from the nearest station to the venue.
\end{abstract}

\ccsdesc[300]{Human-centered computing~Activity centered design}

\keywords{Laser Beam Machining, Laser Cutter, Digital Fabrication, Carving, Leaf, Communication, QR Code}

\begin{teaserfigure}
  \includegraphics[width=\textwidth]{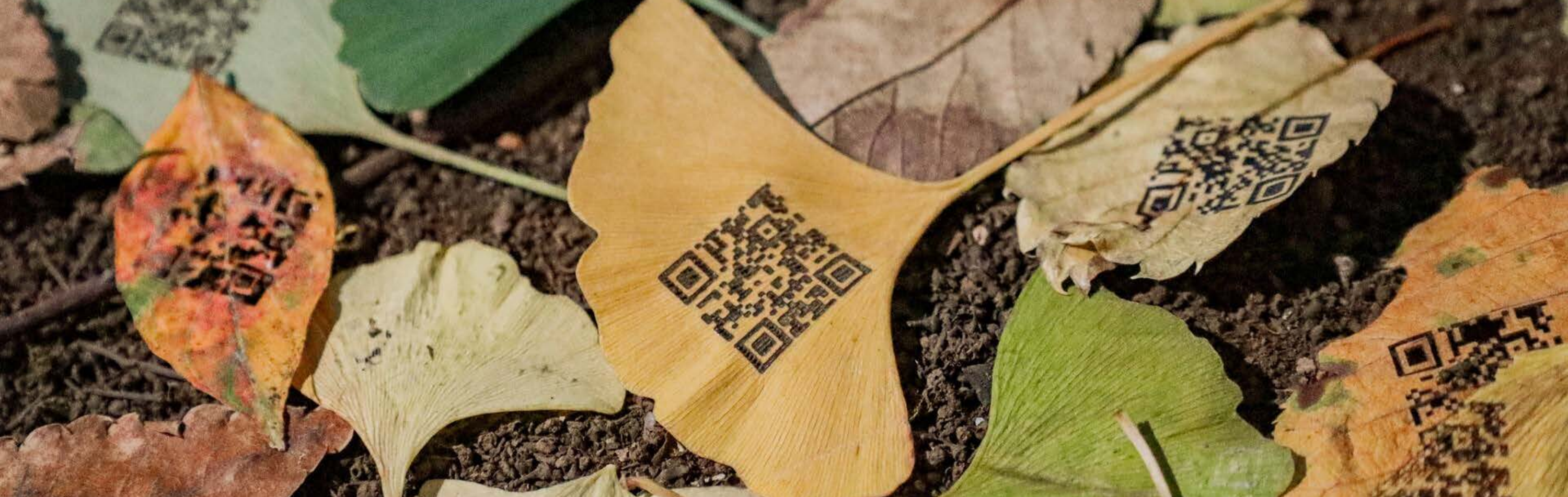}
  \caption{Laser printed leaves. Users can scan the QR code of the leaf to connect to a communal virtual space, where he/she can leave messages to people who would scan the QR code of the leaf itself.}
  \label{fig:teaser}
\end{teaserfigure}


\maketitle

\section{Introduction}
With the widespread use of smartphones and the Internet, we can instantly respond to and communicate messages from family and friends. 
The increasing speed, visibility, and enormity of human connections through these technologies have enriched our daily lives. 
For example, it has become commonplace to interact with our family and friends worldwide 24 hours a day \cite{liu2022, li2022}.
While these technologies have strengthened existing relationships, they make fragile relationships with strangers around us weaker and less visible; in fact, they have made these indirect relationships less visible by facilitating direct communication.
Even though you are unconscious, you share the roads and places around you with many unknown neighbors and creatures. 
Realizing these connections may cultivate social consciousness, thus, we propose the system {\it Ichiyo} for people to focus on those fragile and transient relationships with neighbors around them that are not directly connected or known.
This system is inspired by the words ``Even a chance acquaintance is decreed by destiny.'' in Japanese saying, and {\it Ichiyo} means a leaf in Japanese.
We use leaves as a medium to realize indirect communication between people on the road. 
This is not only because leaves are natural on the road, but are also characterized by their `fragility,' as they can easily be torn by the wind, which is consistent with the ephemeral interaction method we have focused on in this study. 
As there is a limit to the amount of information that can be physically written or carved on a leaf, we decided to virtually expand the amount of information by engraving QR codes on the leaf~\cite{qr_site}. 
Furthermore, we made it possible for users to read the messages of their predecessors who picked up the leaf through the QR code and to leave messages. 
In the production of printing QR codes on leaves, we tried off-the-shelf laser cutters to engrave the QR codes onto the leaves so that there is no chemical or impure pollution to the environment and need to collect the leaves. 
Unlike the method of printing on the leaf surface with ink, this method has less impact on the environment and is less costly; when printing with ink, even a small amount of substances that are not original in nature are scattered.
We have succeeded in producing leaves so that users can scan their QR codes by discovering the moderate imprinting settings and how they should be stored in order not to lose the amount of moisture in the leaves.

Our contributions are presented below:
\begin{itemize}
    \item We proposed a new method of interaction with others: an environmentally friendly method of leaving indirect messages for `fragile' and `transient' interaction in the neighborhood.
    \item We proved that QR codes can be engraved on leaves by off-the-shelf laser cutters and are readable by smartphones.
    \item We demonstrated the appropriate power, speed, and resolution range of a laser cutter when ruling leaves.
\end{itemize}

\section{Methodology}
Here we present the process that led to the completion of {\it Ichiyo}. 


\subsection{Material and Procedure}
For the leaf media engraved with QR codes using a laser cutter, we primarily utilized ginkgo leaves as well as red oak, zelkova, camphor tree, osmanthus, and Japanese chinquapin. To avoid disturbing the ecosystem, we picked up all the leaves that had fallen from the vicinity of the exhibition area. A laser cutter was then used to mark the surface of the leaves and embed a QR code.

\subsection{Implementation}
In this section, we explain the hardware and software aspects of the fabrication.

\begin{table}[]
\caption{Laser cutters' Power, speed, and resolution setting.}
\begin{tabular}{llll}
\hline
                 & Power (\%) & Speed (\%) & Resolution \\ \hline
VLS 6.75--75W    & 7 (=5.25W)     & 30         & 500        \\ \hline
VLS 2.30 DT--30W & 17.5 (=5.25W)      & 30         & 500        \\ \hline
\end{tabular}
\label{tab:setting}
\end{table}

\begin{figure}
    \centering
    \includegraphics[width=0.5\linewidth]{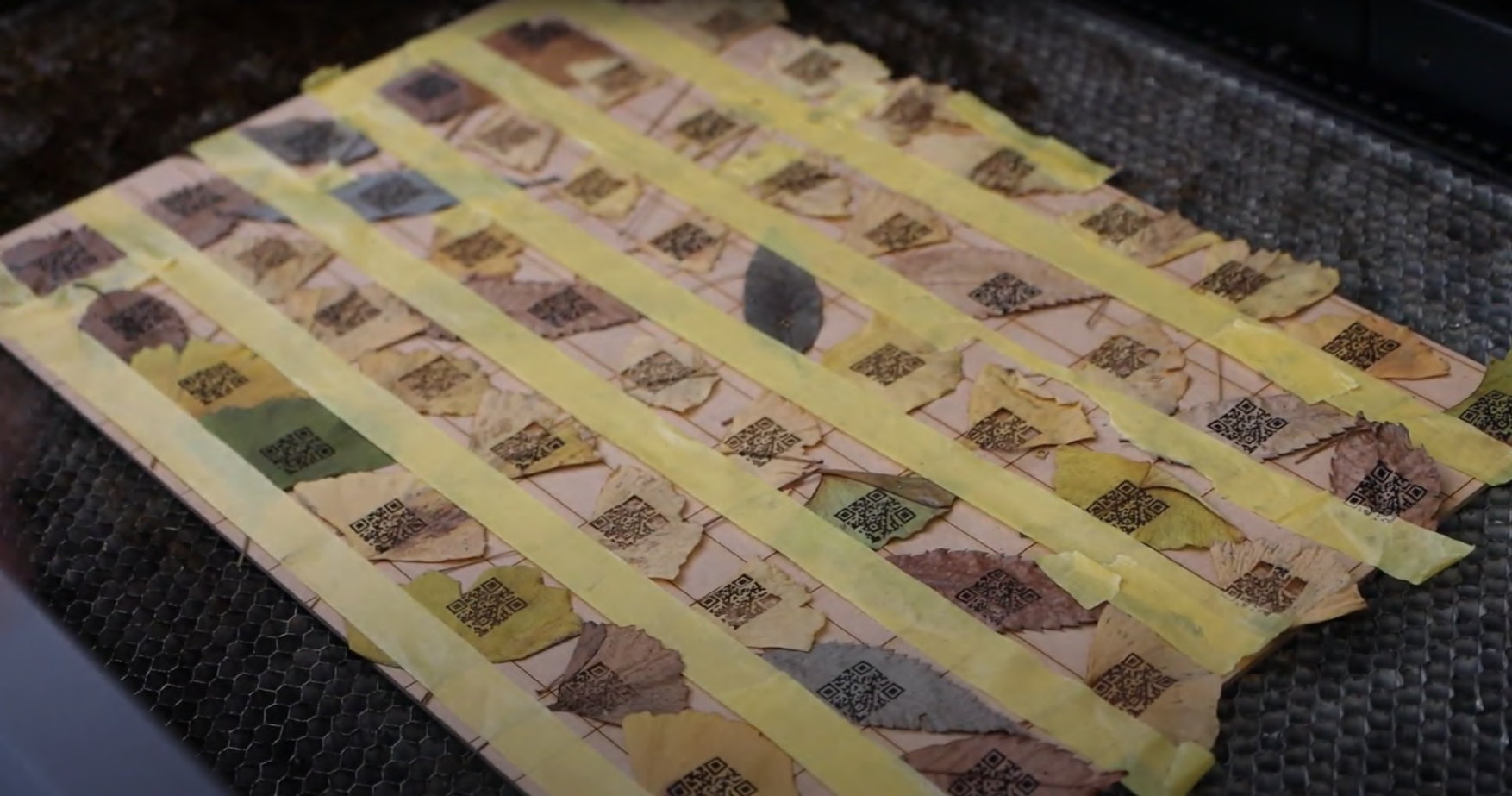}
    \caption{Leaves fixed to the plywood in a grid pattern using curing tape. We engraved QR codes of 20 mm to 30 mm, depending on the size of the leaves.}
    \label{fig:printing}
\end{figure}

\subsubsection{Hardware}
We describe the procedure for printing a single leaf. 
We used the VLS 6.75--75W and VLS 2.30 DT--30W (Universal Laser Systems, Inc.) laser cutter for engraving. 
The power, speed, and resolution are given in Table. \ref{tab:setting}. 
These values were tuned based on our empirical rules coming from massive trial and error. 
Although there were individual differences in leaf thickness and a wide range of suitable settings for outputs, the above setting was applicable to most kinds of leaf and almost all thicknesses.
The focal point was adjusted to suit the plywood. 
The leaves were fixed to the plywood in a grid pattern using curing tape, and QR codes of 20 mm to 30 mm were printed as shown in Fig. \ref{fig:printing}.
We consider QR codes the most accessible means for people to leave messages on a web application.
We then wrapped the leaves in newspaper to dehumidify them, flatten them, and stored them in an enclosed package. 
Although the leaves were flattened and fixed using tape, some leaves were uneven and could not be read. 
In addition, because the leaves were manually placed, they were sometimes misaligned with the QR code.
Therefore, leaves with readable QR codes were produced at a rate of around 18 per hour.
As for the type of QR code, the number of visitors to the exhibition was limited, so the probability of picking up a leaf with the same destination was considered, and the number of destinations was limited to four so that active communication could occur to a certain extent.

\begin{figure}
    \centering
    \includegraphics[width=0.45\linewidth]{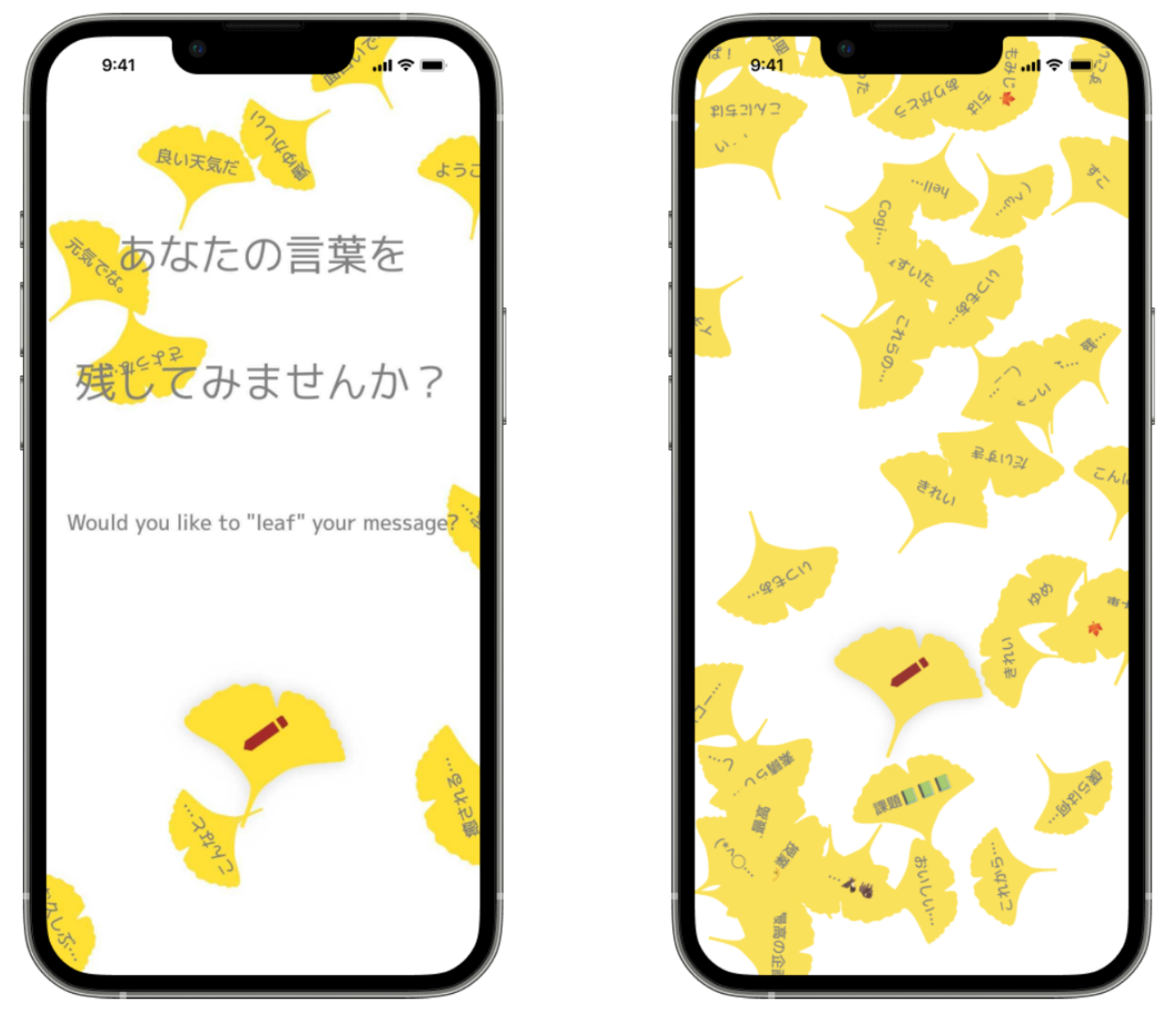}
    \caption{User interface of the web application part of {\it Ichiyo}.}
    \label{fig:ui_ichiyo}
\end{figure}

\subsubsection{Software}
Here, we describe a transition point from a QR code. 
We developed a web application that allowed users to leave their own comments and read the messages of their predecessors who have read the leaves and deployed them at the destination of the URL of the QR codes. 
The user interface of the web application is shown in Fig. \ref{fig:ui_ichiyo}.

\section{Feedbacks}
We exhibited our system {\it Ichiyo} at the public art exhibition ``iii Exhibition 2022'' held at the University of Tokyo, where 1289 people visited our booth. 
We got feedback, especially on the usage of QR codes on natural objects. 
Each visitor mostly imagined what he/she would do with this novel technology.

\section{Application and Future Work}
There are four topics involving application examples and future work.
First, we should test this system on a social scale. 
It is necessary to determine whether this can be used to measure fragile and transient communication with neighbors by communicating using our system and to investigate the results of such a measurement.
Second, the system's potential for other applications should be explored.
Although this system facilitated communication between strangers, the technology of printing QR codes on natural objects can also be used for other applications (e.g., advertising). 
Therefore, it would be a good direction in the future to experiment with QR code technology on natural objects for various applications, such as advertising media.
In addition, more detailed research on reading QR codes on natural objects is necessary. 
For example, it would be of great value to investigate the relationship between reading accuracy and individual inconsistency (e.g., moisture level, type of leaves, and thickness).
Finally, further improvements in the efficiency of the ruling technology should also be considered in the future. 
For example, it is possible to investigate the appropriate output of the laser cutter so that it can be more generally expressed in mathematical formulae or to incorporate a mechanism that can recognize and adjust the position on the leaf at which the QR code is printed and focused when printing QR codes.

\begin{acks}
We would like to express our gratitude to the project advisors and the exhibition owners: Prof. Takeshi Naemura, Prof. Yasuaki Kakehi, Prof. Hidenori Watanabe, Prof. Shohei Takei, Prof. Koki Sone, and Prof. Wataru Date. Their patient, constructive, and inspiring guidance during the design and development process greatly helped us. We also want to sincerely thank the staff at T-BOX~\cite{tbox} who taught us how to manipulate laser cutters and let us use them and the workspace.
We would like to thank Prof. Jun Rekimoto for reviewing our paper and helping us brush up on this paper.
\end{acks}

\bibliographystyle{ACM-Reference-Format}
\bibliography{sample-authordraft}










\end{document}